\documentclass[12pt,a4]{article}
\usepackage{epsfig}
\usepackage{setspace}

\title{Comments on the Refractive Index of Tin Sulphide Nano-crystalline Thin 
Films}

\begin{small}
\author{Amit Jakhar, Ashu Jamdagni, Taruna
Verma, Vibhav Shukla,\\ 
Priyal Jain, Nidhi Sinha and P.Arun\\
Department of Electronics, S.G.T.B. Khalsa College, \\
University of Delhi, Delhi 110007, India\\ \\
\& \\ \\
Ayushi Bakshi\\
Department of Physics, Hansraj College, \\
University of Delhi, Delhi 110007, 
India
}
\end{small}

\begin{document}
\maketitle

\begin{abstract}
The refractive indices of nano-crystalline thin films of Tin (IV) Sulphide 
(SnS) were investigated here. The experimental data conformed well with the 
single oscillator model for refractive indices. Based on the this, we explain 
the increasing trend of refractive index to the improvement in crystal 
ordering with increasing grain size.
\end{abstract}


\newpage
\section{Introduction}
Tin Sulphide (SnS) is a semiconducting material with layered structure 
\cite{b.a} that has attracted much attention in the past due to its potential 
application in optical data storage \cite{b.b} and as an absorbing layer in 
solar cells \cite{b.c,b.d}. Hence, various techniques have been looked into 
to fabricate its good quality thin films \cite{b.e,b.f,b.g,b.h,b.i}. While 
much has been discussed about SnS thin film's band-gap and its strong 
dependence on fabrication method, relatively less has been done to study the 
material's refractive index under various conditions. In this manuscript, we 
discuss the variation of SnS nano-crystalline film's refractive index with 
grain size. 

\section{Experimental}
Thin films of  tin sulphide were deposited using thermal evaporation 
technique on flat glass substrates. The depositions were carried out at 
room temperature in a vacuum coating unit (Hind Hi-Vac 12A4D) with vacuum 
better than ${\rm 4 \times 10^{-5}}$~Torr. The deposition rate was maintained 
constant at 2.7~\AA/s. The deposition current and time were controlled to 
fabricate films of varying thicknesses. A portion of these films were then 
annealed at two different temperatures, 373 and 473K, for 30 minutes under 
vacuum of ${\rm 10^{-3}}$~Torr. Thickness of the films were measured using 
Dektak surface profiler (150). The influence of film thickness and annealing 
on the structure of the films were checked using X-ray Diffractometer (Bruker 
D8 X-ray Diffractometer) operating at the 40 KV and 40 mA with 
${\rm CuK\alpha}$ radiation (λ=1.5406~\AA) and Transmission Electron 
Microscopy (Technai T30U Twin). The optical absorption and transmission 
spectra of the films were recorded as a function of wavelength using an 
UV-Vis Double Beam Spectrophotometer (Systronics 2202) over the range 
300-1100~nm with unpolarized light incident normal to the film surface. 

\section{Results and Discussion}
\subsection{The Structural and Morphological Analysis}
\begin{figure}
\begin{center}
\epsfig{file=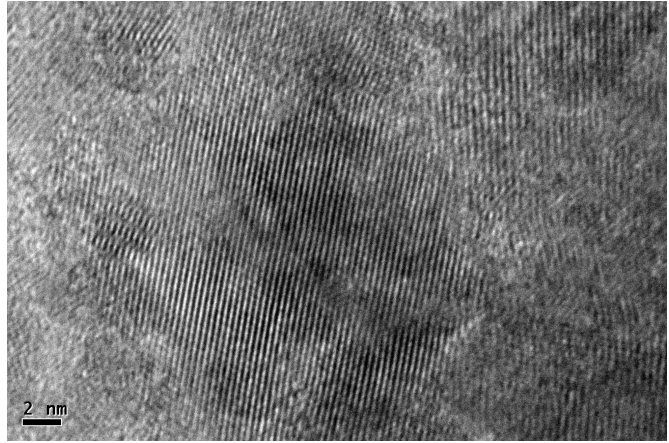, width=3.45in, angle=-0}
\end{center}
\caption{Parallel (040) planes are clearly seen in HRTEM image. }
\label{fig.1}
\end{figure}

As grown and annealed (at 373 K and 473 K) SnS thin films of different 
thicknesses (namely 270, 480, 600, 630 and 900~nm) were used in this study. 
The films were found to be uniform and dark brown in color. There was 
negligible or no change in the color of the films upon annealing. SnS is 
known to have a layered structure as mentioned in the introduction. The High 
Resolution Transmission Electron Microscope (HRTEM) image of a 480~nm thick 
SnS film shown in fig.~\ref{fig.1} clearly shows that the film retains the 
layered structure of its bulk. The crystallographic structures of the as 
grown and annealed SnS films were analysed by X-ray Diffraction.  

\begin{figure}
\begin{center}
\epsfig{file=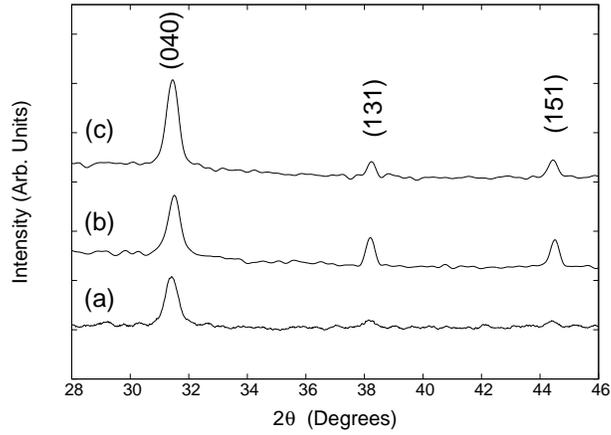, width=2.35in, angle=-90}
\end{center}
\caption{X-ray diffraction pattern of a 480~nm thick SnS (a) as grown film. 
Pattern  (b) and (c) are those of same film after  annealing at  373~K and 
473~K respectively.}
\label{fig.2}
\end{figure}

Fig.~\ref{fig.2} shows a representative X-ray diffractogram of as grown and 
annealed SnS films of thickness 480 nm. Without exception, the diffraction 
patterns exhibits (fig~2) peaks at (${\rm 2\theta}$)=31.1$^o$, 38.5$^o$ and 
43.6$^o$ corresponding to the (040), (131) and (200) planes which matched 
well with the orthorhombic structure reported in ASTM card No. 83-1758. The 
patterns also show that diffraction peaks become more sharp as the annealing 
temperature increases, possibly implying that the films become more 
crystalline on annealing. While improvement in ordering with annealing is 
evident, we found that there was no variation in structure with annealing. 
The lattice parameters of the unit cell (`a', `b', and `c') were estimated by 
using the positions of the (040),  (200) and (131) XRD peaks. Experimentally 
obtained values of  `a', `b' and `c' of the  samples were found to be 
${\rm 4.07\pm0.002}$, ${\rm 11.36\pm0.01}$ and ${\rm 4.45\pm0.002~\AA}$ 
respectively. The as grown samples of thicknesses 270 and 480~nm were 
strained, i.e. their lattice parameter (`a') was slightly different. However, 
this residual stress relaxed on annealing. Only samples with the same lattice 
parameters are considered here.

The average grain size of the films were calculated from the Full Width at 
Half Maxima (FWHM) of the XRD peaks using the Scherrer’s formula 
\cite{b.i,b.j,b.j1}
\begin{equation}
\label{eq.1}
r={0.9 \lambda \over \beta cos \theta}
\end{equation}
where `r' is the grain size, ${\rm \beta}$ is the FWHM, ${\rm \theta}$ and 
${\rm \lambda}$ have their usual meaning. The grain sizes were found to vary 
from 11 to 25~nm depending on the film thickness and annealing temperature 
(details of which we have reported elsewhere \cite{b.k}).

\subsection{Optical Analysis}
\begin{figure}
\begin{center}
\epsfig{file=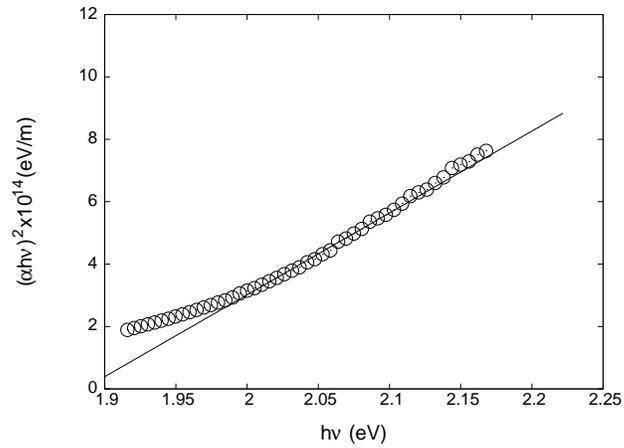, width=2.35in, angle=-90}
\end{center}
\caption{Plot shows the variation of  ${\rm (\alpha h\nu)^2}$ with photon 
energy (${\rm h\nu)}$. The linear region when extrapolated on to the `X'-axis 
gives the ``allowed-direct transition'' optical band-gap.}
\label{fig.3}
\end{figure}

From the above analysis it is clear that the films under investigation thus 
differed only in grain size. Hence, this presented us the opportunity to 
study the variation of optical properties of SnS as a function of grain size 
that too in the nano regime. The optical property of a material is represented 
by its band gap and its refractive index. We use the standard Tauc method 
\cite{b.m} to investigate the band gap of SnS films. Fig.~\ref{fig.3} shows 
a ${\rm (\alpha h\nu)^2}$ vs ${\rm h\nu}$ plot for a 480~nm thick film where 
`${\rm \alpha}$' is the absorption coefficient and ${\rm h\nu}$ the photon 
energy. The band gap of the film was estimated by extrapolating the linear 
part of the plot to the ${\rm h\nu}$ axis.

\begin{figure}
\begin{center}
\epsfig{file=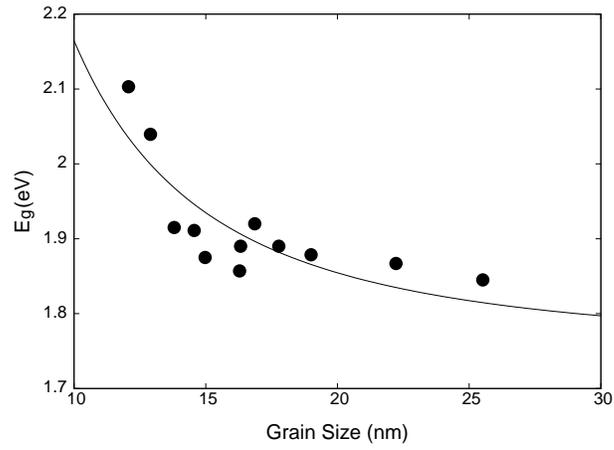, width=2.35in, angle=-90}
\end{center}
\caption{Variation of band-gap with grain size.}
\label{fig.4}
\end{figure}

The variation in the band gap with grain size of the films is shown in 
fig.~\ref{fig.4}. The trend in variation fits to 
\begin{equation}
\label{eq.2}
E=E_g(bulk)+{A \over r^2}
\end{equation}
where Eg(bulk) is the band gap of SnS in bulk and `A' a constant relating to 
the effective mass of the charge carriers in SnS. This result is consistent 
with properties induced by quantum confinement of electrons \cite{b.n}. Again 
this result confirms that SnS grains of 11-25 nm are in the nano-regime. 

Many studies investigating the dependence of refractive index on the thickness 
\cite{b.o,b.p}, annealing temperature \cite{b.q} and substrate temperature 
\cite{b.r} have been reported in the literature but no work on its dependence 
on grain size has been reported so far. Thus, it will be fruitful to 
investigate the grain size dependence of the refractive index of the films. 
The transmission spectra of the films exhibit interference fringes within the 
wavelength range 300-1100~nm.

\begin{figure}
\begin{center}
\epsfig{file=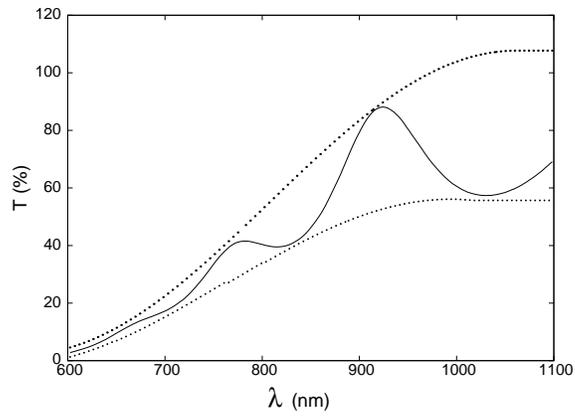, width=2.25in, angle=-90}
\end{center}
\caption{The transmission spectra of a SnS thin film with envolopes used to 
calculate the film's refractive index.}
\label{fig.5}
\end{figure}

The envelope shown in the fig.~\ref{fig.5} were drawn connecting the maximas 
and minimas of the interference peaks. Such envelopes were used to estimate 
the refractive index of the films using Swanepoel’s method \cite{b.s}. 
Swanepoel method gives the refractive index in the transparent and near band 
edge region as 
\begin{equation}
\label{eq.3}
n=\sqrt{N_1+\sqrt{(N_1^2-s^2)}}
\end{equation}
where
\begin{equation}
\label{eq.4}
N_1={2s \over T_m}+{s^2+1 \over 2}
\end{equation}
and ${\rm T_m}$ is the minimas of the interference fringes and `s' is the 
refractive index of the glass substrate. The values of the refractive index 
(1.5-2.2)  are well with in the range found in the literature \cite{b.t}. 
Evaluated results show a decrease in the refractive index `n' with increasing 
wavelength due to the normal dispersion behavior of SnS \cite{b.u}. Also, the 
values of the refractive index of the films deposited at room temperature were 
found to be less than those annealed at 373 and 473~K.

\begin{figure}
\begin{center}
\epsfig{file=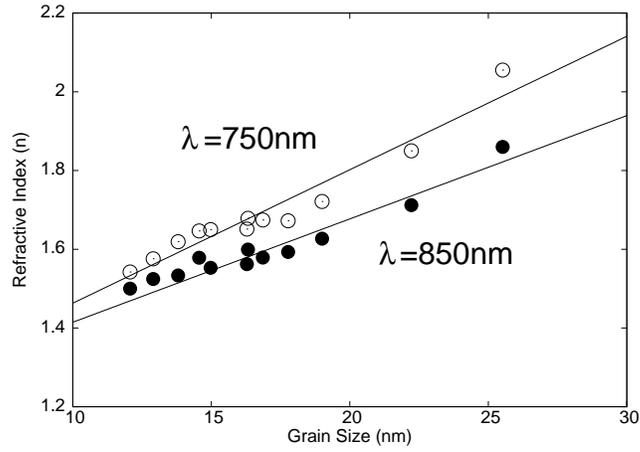, width=2.35in, angle=-90}
\end{center}
\caption{Variation of refractive index with grain size for two different 
wavelengths.}
\label{fig.6}
\end{figure}

\begin{figure}
\begin{center}
\epsfig{file=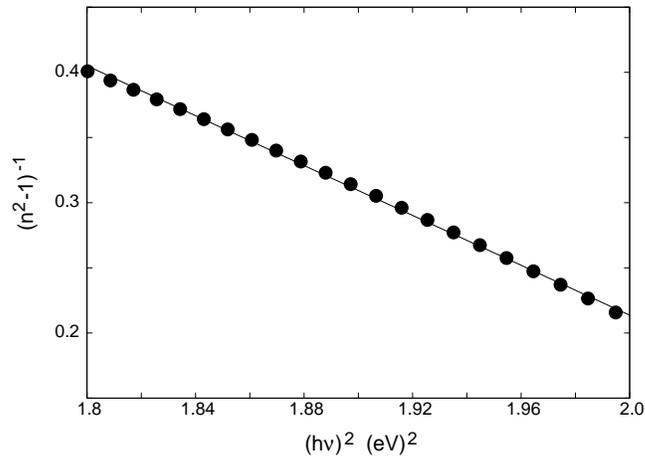, width=2.5in, angle=-90}
\end{center}
\caption{Representative plot used to estimate ${\rm E_o}$ and ${\rm E_d}$.}
\label{fig.7}
\end{figure}

We believe this is a result of improving grain size with annealing. The 
variation of the refractive index with the grain size is shown in 
fig.~\ref{fig.6}. The refractive index `n' of the films follows an increasing 
trend with the grain size. A pertinent question is why would the refractive 
index increase with the grain size? To investigate into this question we fit 
our data to Wemple and DiDomenico (WDD) model \cite{b.v,b.w}. Based on the 
idea that refractive index of the materials in visible region is due to 
electron oscillation excitation between conduction and valence band 
(inter-band transitions in and around the band edge), the refractive index is 
given as: 
\begin{equation}
\label{eq.5}
n^2=1+{E_oE_d \over E_o^2+(h\nu)^2}
\end{equation}

where `${\rm h\nu}$' is the photon energy, ${\rm E_o}$ is the single 
oscillator energy and ${\rm E_d}$ is the dispersion energy. The dispersion 
energy, ${\rm E_d}$, is the strength of transition and gives the strength of 
the oscillator.  These parameters are derived by fitting a linear function in 
${\rm (n^2-1)^{-1}}$ against ${\rm (h\nu)^2}$ plot. Fig.~\ref{fig.7} shows a 
representative plot of ${\rm (n^2-1)^{-1}}$ against ${\rm (h\nu)^2}$ of a 
480~nm film annealed at 373~K. The intercept (c) and slope (m) of the best 
fit line to the experimental data give the value of ${\rm E_o}$ 
(${\rm =\sqrt{c/m}}$) and ${\rm E_d}$ (${\rm =1/ \sqrt{mc}}$).

\begin{figure}
\begin{center}
\epsfig{file=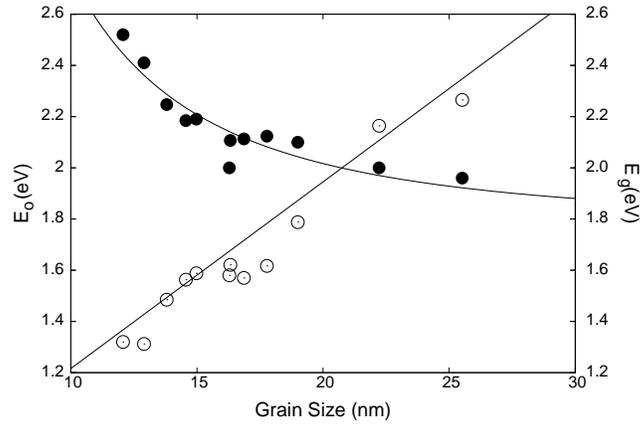, width=2.25in, angle=-90}
\end{center}
\caption{Variation of ${\rm E_o}$ and ${\rm E_d}$, parameters related to the 
Wemple and DiDomenico's single oscillator model, shown with grain size. 
While ${\rm E_o}$ follows the same trend as the material's band-gap, 
${\rm E_d}$ increases with increasing grain size.}
\label{fig.8}
\end{figure}

Fig.~\ref{fig.8} shows the variation of ${\rm E_o}$ and ${\rm E_d}$ with 
increasing grain size. While ${\rm E_o}$ decreases with grain size, 
${\rm E_d}$ increases linearly with grain size. ${\rm E_d}$ is given as 
\begin{equation}
\label{eq.6}
E_d= \beta N_cN_eZ_a
\end{equation}
where ${\rm \beta}$ is the a constant whose value depends on the chemical 
bonding character of the material, ${\rm N_c}$ is the co-ordination number, 
${\rm N_e}$ is the total number of valence electrons per anion and 
${\rm Z_a}$ is the chemical valency of the anion. Literature gives the values 
of ${\rm \beta, N_c, N_e, Z_a}$ as 0.39~eV, 3, 10 and 2 respectively 
\cite{b.x}. Considering the samples are identical (barring for change in 
grain size)  with no changes taking place in the crystal structure, one 
expects ${\rm E_d}$ to remain constant. However, changes is ${\rm E_d}$ have 
been reported before and explained due to increased ordering \cite{b.o}. Our 
samples have exhibited improvement in crystalline nature and increased grain 
size on  annealing, as indicated from the XRD plots. Hence, the trend in
${\rm E_d}$ with grain size is understood.

The observed decrease in the ${\rm E_o}$ with increasing anneal temperatures 
can be attributed to quantum confinement in the films. As mentioned, 
${\rm E_o}$ is the single oscillator energy related to to the material’s band 
gap. We hence, plot ${\rm E_o}$ with respect to ${\rm E_g}$ (figure 9). We 
find ${\rm E_o}$ is directly proportional to ${\rm E_g (\approx 1.91 E_g)}$. 
This is consistent with results reported in literature \cite{b.o}.  It is 
hence natural that the variation of ${\rm E_o}$ with the grain size is 
identical to that of the optical band gap (see figure 6).

\section{Conclusions}
Thin nano-crystalline films of SnS were fabricated on glass substrates kept at room temperature. These films were also annealed. Samples having the same crystalline structure (namely identical lattice parameters) were selected for comparing their refractive indices. Study showed that the film's refractive index is explainable using the single oscillator model and is directly proportional to grain size.   

\subsection*{Acknowledgments}
This work was supported by University of Delhi under its ``Innovation
Projects'' program (SGTB-101). The authors also acknowledge the assistance rendered by Dr. Binay Kumar, Department of Physics and Astrophysics, University of Delhi.

\newpage

\section*{Figure Captions}
\begin{itemize}
\item[1.] Parallel (040) planes are clearly seen in HRTEM image.
\item[2.] X-ray diffraction pattern of a 480~nm thick SnS (a) as grown film. 
Pattern  (b) and (c) are those of same film after  annealing at  373~K and 
473~K respectively.
\item[3.] Plot shows the variation of  ${\rm (\alpha h\nu)^2}$ with photon 
energy (${\rm h\nu)}$. The linear region when extrapolated on to the `X'-axis 
gives the ``allowed-direct transition'' optical band-gap.
\item[4.] Variation of band-gap with grain size.
\item[5.] The transmission spectra of a SnS thin film with envolopes used to 
calculate the film's refractive index.
\item[6.] Variation of refractive index with grain size for two different 
wavelengths.
\item[7.] Representative plot used to estimate ${\rm E_o}$ and ${\rm E_d}$.
\item[8.] Variation of ${\rm E_o}$ and ${\rm E_d}$, parameters related to the 
Wemple and DiDomenico's single oscillator model, shown with grain size. 
While ${\rm E_o}$ follows the same trend as the material's band-gap, 
${\rm E_d}$ increases with increasing grain size.
\item[9.] Variation of ${\rm E_o}$ with ${\rm E_g}$.
\end{itemize}

\newpage

\section*{Figures}

\begin{figure}
\begin{center}
\epsfig{file=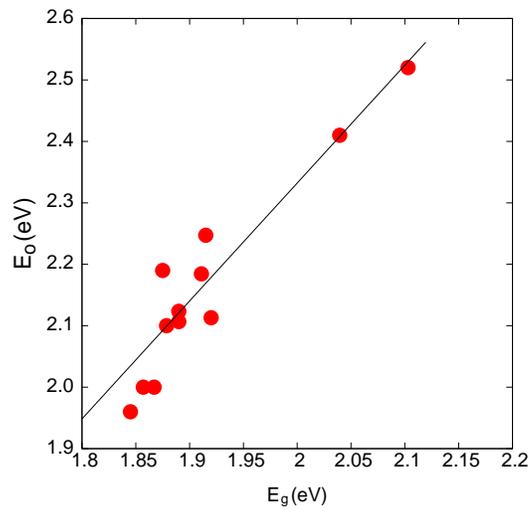, width=2.7in, angle=-90}
\end{center}
\caption{Variation of ${\rm E_o}$ with ${\rm E_g}$.}
\label{fig.9}
\end{figure}

\end{document}